\documentclass[aps,12pt,superscriptaddress,preprint]{revtex4}

\usepackage[dvips]{graphicx}
\usepackage{amsmath,amssymb,amsthm,bm}

\newcommand{\nn}{\nonumber} 
\renewcommand{\epsilon}{\varepsilon} 
\renewcommand{\phi}{\varphi}

\theoremstyle{plain}
\newtheorem*{theorem}{Theorem}
\newtheorem{definition}{Definition}

\begin{document}

\preprint{IUHET-522}

\vspace*{0.75in}

\title{Decomposition of geometric perturbations}

\author{Roman V. Buniy}
\email{roman.buniy@gmail.com}
\affiliation{Physics Department, Indiana University, Bloomington, IN 47405}

\author{Thomas W. Kephart} 
\email{tom.kephart@gmail.com}
\affiliation{Department of Physics and Astronomy, Vanderbilt
University, Nashville, TN 37235}

\date{November 12, 2008}

\begin{abstract}

For an infinitesimal deformation of a Riemannian manifold, we prove
that the scalar, vector, and tensor modes in decompositions of
perturbations of the metric tensor, the scalar curvature, the Ricci
tensor, and the Einstein tensor decouple if and only if the manifold
is Einstein. Four-dimensional space-time satisfying the condition of
the theorem is homogeneous and isotropic. Cosmological applications
are discussed.

\end{abstract}

\maketitle


\section{Introduction}

The theory of small perturbations in general
relativity~\cite{perturbations} is an important tool in contemporary
cosmology. Since the observed universe is almost homogeneous and
isotropic, the unperturbed space-time is usually assumed to have the
Robertson-Walker form. Although this is a good approximation, recent
cosmological observations suggest the necessity of exploring beyond
it. Without the assumptions of homogeneity and isotropy, a variety of
complications arise, one of which is the mixing of various
perturbation modes. Here we explore this issue.
 
Let $M$ be a manifold with the metric tensor $g$. For an infinitesimal
deformation of $M$ with the perturbation of the metric tensor $\delta
g$, the perturbation of the Ricci tensor $\delta R$ is a linear
functional of $\delta g$. It is convenient to decompose $\delta g$
into irreducible parts relative to $g$. This leads to the
decomposition of $\delta R$. In general, the decompositions of $\delta
g$ and $\delta R$ are coupled. (The precise definition is given in the
next section.) This is due to the fact that local preferred directions
specified by $g$ are only a subset of directions specified by the
unperturbed Ricci tensor $R$ and the unperturbed Weyl tensor $W$. The
purpose of this note is to find necessary and sufficient conditions on
$g$ for which the decompositions of $\delta g$ and $\delta R$
decouple, and its implications for cosmology.

A manifold of constant curvature is an example for which the
decompositions decouple. Manifolds of constant curvature are
classified and are used as model spaces for various constructions in
differential geometry. For such a manifold, $R=\tfrac{1}{n}Sg$ and
$W=0$, where $n=\dim{M}$ and the constant $S$ is the scalar curvature
of $M$. As a result, the only preferred directions in this case are
those specified by $g$. This example suggests that the key property
needed for the decompositions of $\delta g$ and $\delta R$ to decouple
is an appropriate generalization of a manifold of constant curvature.

A natural generalization of a manifold of constant curvature is an
Einstein manifold. For such a manifold, $R=\tfrac{1}{n}Sg$, where $S$
is a constant. Properties of $W$ imply that it can enter a relation
between the terms of the tensor type only, which prevents $W$ from
coupling the decompositions. In two and three dimensions, the only
tensor with the required properties of the Weyl tensor is the zero
tensor. This implies that a manifold of dimension two or three is
Einstein if and only if it is a manifold of constant curvature.

As our main result, we prove a theorem which states, in particular,
that the decompositions of $\delta g$ and $\delta R$ decouple if and
only if $M$ is an Einstein manifold.

The above considerations have application to geometric theories of
gravity. In such a theory, the action functional is $I+J$, where $I$
and $J$ are the geometric and matter parts of the action,
respectively. The equation of motion is $G+T=0$, where $G$ and $T$ are
the variational derivatives of $I$ and $J$, respectively, with respect
to $g$. $G$ is a generalization of the Einstein tensor and $T$ is the
energy-momentum tensor. In this case, we are interested in the
decompositions of $\delta g$ and $\delta G$. When the decompositions
decouple, the equation of motion reduces to separate equations for
each irreducible part, which usually simplifies their solutions. When
the decompositions are coupled, it is usually harder to solve the
equations, but in such cases there are interesting classes of
solutions where perturbations of one type in $\delta T$ leads to
perturbations of multiple types in $\delta g$.

\section{Decomposition theorem}

Let $M$ be an $n$-dimensional manifold with metric $g$ and connection
$\nabla$.
\begin{definition}
  Let $h$ be a symmetric tensor of type $(0,2)$ on $(M,g,\nabla)$. The
  set $(\phi,\psi,\theta,\omega)$ is called the decomposition of the
  tensor $h$ if
  \begin{align}
    h_{ij} &=\tfrac{1}{n}\phi g_{ij} +2(\nabla_i\nabla_j -\tfrac{1}{n}
    g_{ij}\Delta)\psi +\nabla_i \theta_j +\nabla_j \theta_i
    +\omega_{ij},\label{decomposition}
  \end{align}
where
\begin{align}
  \nabla^i \theta_i=0, \quad {\omega^i}_i=0, \quad \nabla^j
  \omega_{ij}=0.
\end{align}
\end{definition}
We obtain expression~\eqref{decomposition} from the
decomposition~\cite{York}
\begin{align}
  h_{ij} &=\tfrac{1}{n}\phi g_{ij} +\nabla_i \xi_j +\nabla_j \xi_i
  -\tfrac{2}{n}g_{ij}\nabla_k\xi^k +\omega_{ij},
\end{align}
where $\phi$ is a scalar, $\xi$ is a vector, and $\omega$ is a
symmetric, traceless, and transverse tensor of type $(0,2)$. It
follows that $\phi={h^i}_i$, and the condition $\nabla^j
\omega_{ij}=0$ then gives
\begin{align}
  \nabla^j\nabla_i\xi_j +\Delta\xi_i
  -\tfrac{2}{n}\nabla_i\nabla^j\xi_j &=\nabla^j(h_{ij}
  -\tfrac{1}{n}\phi g_{ij}).\label{killing}
\end{align}
The solution $\xi$ of Eq.~\eqref{killing} is unique up to a conformal
Killing vector $\eta$, which satisfies the equation
\begin{align}
  \nabla^j\nabla_i\eta_j +\Delta\eta_i
  -\tfrac{2}{n}\nabla_i\nabla^j\eta_j &=0.
\end{align}

Further decomposing $\xi_i=\nabla_i\psi+\theta_i$, where $\psi$ is a
scalar and $\theta$ is a transverse vector, we arrive at
decomposition~\eqref{decomposition}. It follows that the decomposition
$(\phi,\psi,\theta,\omega)$ of $h$ is unique and irreducible. For a
compact Riemannian $(M,g,\nabla)$, one can further prove that the
decomposition is orthogonal. The proof is a trivial extension of the
proof given in Ref.~\cite{York}.
 
\begin{definition}
  Let $f$ be a scalar. Let $h$ and $h'$ be symmetric tensors of type
  $(0,2)$ with the decompositions $(\phi,\psi,\theta,\omega)$ and
  $(\phi',\psi',\theta',\omega')$, respectively. We say that the
  decompositions of $h$ and $f$ decouple if $f$ does not depend on
  $\theta$ and $\omega$. We say that the decompositions $h$ and $h'$
  decouple if $\phi'$ and $\psi'$ do not depend on $\theta$ and
  $\omega$, $\theta'$ does not depend on $\phi$, $\psi$, and $\omega$,
  and $\omega'$ does not depend on $\phi$, $\psi$, and $\theta$.
\end{definition}

For an infinitesimal deformation of $M$ with the metric perturbation
$\delta g$, the perturbations of the scalar curvature, the Ricci
tensor, and the Einstein tensor are
\begin{align}
\delta S &=(\nabla^i\nabla^j -g^{ij}\Delta -R^{ij})\delta g_{ij},
\\ \delta R_{ij} &=\tfrac{1}{2} (\nabla^k\nabla_i \delta g_{kj}
+\nabla^k\nabla_j \delta g_{ki} -\Delta \delta g_{ij}
-g^{kl}\nabla_i\nabla_j \delta g_{kl} ), \\ \delta G_{ij}
&=\tfrac{1}{2} \bigl( \nabla^k\nabla_i \delta g_{kj} +\nabla^k\nabla_j
\delta g_{ki} -(\Delta+S)\delta g_{ij} \bigr) \nn \\ &+\tfrac{1}{2}
\bigl( -g^{kl}\nabla_i\nabla_j +g_{ij}(-\nabla^k\nabla^l +g^{kl}\Delta
+R^{kl}) \bigr) \delta g_{kl}.
\end{align}
\begin{theorem}
  Let $\delta g$, $\delta S$, $\delta R$, and $\delta G$ be
  infinitesimal perturbations of the metric tensor, the scalar
  curvature, the Ricci tensor, and the Einstein tensor, respectively,
  on a manifold $M$. The following statements are equivalent:
  \begin{enumerate}
  \item $M$ is an Einstein manifold.
  \item The decompositions of $\delta g$ and $\delta S$ decouple.
  \item The decompositions of $\delta g$ and $\delta R$ decouple.
  \item The decompositions of $\delta g$ and $\delta G$ decouple.
  \end{enumerate}
\end{theorem}
\begin{proof}
Let $(\phi,\psi,\theta,\omega)$ be the decomposition of $\delta g$. We
find
\begin{align}
  \delta S &=\tfrac{1}{n}\bigl( (1-n)\Delta -S \bigr)\phi
  +\tfrac{2}{n}\bigl( (n-1)\Delta\Delta +S\Delta +n(\nabla^i
  R_{ij})\nabla^j \bigr)\psi \nn \\ &+2(\nabla^i R_{ij})\theta^j
  -R_{ij}\omega^{ij}, \\ \delta R_{ij} &=\tfrac{1}{2n} \bigl(
  (2-n)\nabla_i\nabla_j -g_{ij}\Delta \bigr)\phi \nn
  \\ &+((1-\tfrac{2}{n})\nabla_i\nabla_j\Delta
  +\tfrac{1}{n}g_{ij}\Delta\Delta +R_{ik}\nabla_j\nabla^k
  +R_{jk}\nabla_i\nabla^k +(\nabla_k R_{ij})\nabla^k)\psi \nn
  \\ &+R_{ik}\nabla_j\theta^k +R_{jk}\nabla_i\theta^k +(\nabla_k
  R_{ij})\theta^k \nn \\ &-\tfrac{1}{2}\Delta\omega_{ij}
  +\tfrac{1}{2}R_{ik}{\omega_j}^k +\tfrac{1}{2}R_{jk}{\omega_i}^k
  -P_{ikjl}\omega^{kl}, \\ \delta G_{ij} &=\tfrac{1}{2n}
  (2-n)(\nabla_i\nabla_j -g_{ij}\Delta \bigr)\phi \nn \\ &+ \bigl(
  \tfrac{1}{n}(n-2)(\nabla_i\nabla_j-g_{ij}\Delta)\Delta
  -S\nabla_i\nabla_j -\tfrac{1}{2}g_{ij}(\nabla_k S)\nabla^k \nn
  \\ &+R_{ik}\nabla_j\nabla^k +R_{jk}\nabla_i\nabla^k +(\nabla_k
  R_{ij})\nabla^k \bigr)\psi \nn \\ &+R_{ik}\nabla_j\theta^k
  +R_{jk}\nabla_i\theta^k +(\nabla_k R_{ij})\theta^k
  -\tfrac{1}{2}S(\nabla_i\theta_j +\nabla_j\theta_i)
  -\tfrac{1}{2}g_{ij}(\nabla_k S)\theta^k \nn
  \\ &-\tfrac{1}{2}(\Delta+S)\omega_{ij}
  +\tfrac{1}{2}R_{ik}{\omega_j}^k +\tfrac{1}{2}R_{jk}{\omega_i}^k
  +\tfrac{1}{2}g_{ij}R_{kl}\omega^{kl} -P_{ikjl}\omega^{kl},
\end{align}
where $P$ is the Riemann tensor of type $(0,4)$. Since $\theta$ is an
arbitrary transverse vector and $\omega$ is an arbitrary symmetric,
traceless, and transverse tensor, it follows that $\delta S$ does not
depend on $\theta$ and $\omega$ if and only if $R=\tfrac{1}{n}Sg$,
where $S$ is a constant. In this case we find
\begin{align}
  \delta S &=\tfrac{1}{n}\bigl( (1-n)\Delta -S \bigr)\phi
  +\tfrac{2}{n}\bigl( (n-1)\Delta +S \bigr)\Delta\psi.
\end{align}
This proves that statements $1$ and $2$ are equivalent.

Let $(\phi',\psi',\theta',\omega')$ be the decomposition of $\delta
R$. From $\phi'=g^{ij}\delta R_{ij}$ we find
\begin{align}
  \phi' &=\tfrac{1}{n}(1-n)\Delta\phi +\tfrac{2}{n}((n-1)\Delta\Delta
  +n\nabla^i R_{ij}\nabla^j)\psi +2\nabla^i R_{ij}\theta^j.
\end{align}
It follows that $\phi'$ does not depend on $\theta$ if and only if
$R=\tfrac{1}{n}Sg$, where $S$ is a constant. In this case we find
\begin{align}
  \phi' &=\tfrac{1}{n}(1-n)\Delta\phi +\tfrac{2}{n}((n-1)\Delta
  +S)\Delta\psi, \\ \psi' &=\tfrac{1}{4n}(2-n)\phi
  +\tfrac{1}{2n}\bigl( (n-2)\Delta +2S \bigr)\psi, \\ \theta'_i
  &=\tfrac{1}{n}S\theta_i, \\ \omega'_{ij} &=\bigl(
  -\tfrac{1}{2}\Delta +\tfrac{1}{n-1}S \bigr)\omega_{ij}
  -W_{ikjl}\omega^{kl}.
\end{align}
This proves that statements $1$ and $3$ are equivalent.

Let $(\phi'',\psi'',\theta'',\omega'')$ be the decomposition of
$\delta G$. From $\phi''=g^{ij}\delta G_{ij}$ we find
\begin{align}
  \phi'' &=\tfrac{1}{2n}(n-1)(n-2)\Delta\phi +\bigl(
  \tfrac{1}{n}(n-1)(2-n)\Delta\Delta -S\Delta
  +(1-\tfrac{n}{2})(\nabla_i S)\nabla^i \nn
  \\ &+2R_{ij}\nabla^i\nabla^j \bigr)\psi +2R_{ij}\nabla^i\theta^j
  +(1-\tfrac{n}{2})(\nabla_i S)\theta^i
  +\tfrac{n}{2}R_{ij}\omega^{ij}.
\end{align}
It follows that $\phi''$ does not depend on $\theta$ and $\omega$ if
and only if $R=\tfrac{1}{n}Sg$, where $S$ is a constant. In this case
we find
\begin{align}
  \phi'' &=\tfrac{1}{2n}(n-1)(n-2)\Delta\phi +\tfrac{1}{n}(2-n)\bigl(
  (n-1)\Delta +S \bigr)\Delta\psi, \\ \psi'' &=\tfrac{1}{4n}(2-n)\phi
  +\tfrac{1}{2n}(n-2)(\Delta -S)\psi, \\ \theta''_i &=
  \tfrac{1}{2n}(2-n)S\theta_i, \\ \omega''_{ij} &= \bigl(
  -\tfrac{1}{2}\Delta +\tfrac{3-n}{2(n-1)}S \bigr)\omega_{ij}
  -W_{ikjl}\omega^{kl}.
\end{align}
This proves that statements $1$ and $4$ are equivalent and thus
concludes the proof of the theorem.
\end{proof}

\section{Application}

We now apply our theorem to the theory of small perturbations in
general relativity~\cite{perturbations}. We consider an
$(n+1)$-dimensional pseudo-Riemannian space-time $(M',g')$ of
signature $(1,n)$ as locally foliated by $n$-dimensional Riemannian
spaces $(M(t),g(t))$ which depend on the temporal coordinate $t$ as a
parameter. We use Gaussian normal coordinates on $M'$, so that
$g'_{00}=-1$, $g'_{0i}=0$, $g'_{ij}=g_{ij}(t)$, where the index $0$
refers to the coordinate $t$. We can always choose these coordinates
in such a way that they remain locally Gaussian normal after an
infinitesimal deformation of $M'$, so that $\delta g'_{00}=0$, $\delta
g'_{0i}=0$, $\delta g'_{ij}=\delta g_{ij}(t)$.

By the theorem, the decompositions of $\delta g(t)$ and $\delta G(t)$
decouple if and only if $M(t)$ is an Einstein manifold. Consider now
the case $n=3$, which is particularly important in cosmology. Since
$W(t)=0$ in such a case, it follows that $M(t)$ is an Einstein
manifold if and only if it is a manifold of constant curvature. If
$M(t)$ is a manifold of constant curvature for each $t$, then $M'$ is
a homogeneous, isotropic space-time. In such a case, the metric $g'$
is essentially uniquely specified by the homogeneity and isotropy
conditions, and depends on the scalar curvature $S(t)$. We thus arrive
at the standard Robertson-Walker cosmological model, for which
decoupling of the decompositions is well-known. The theorem implies,
in particular, that such a space-time is the only case in which the
decompositions of $\delta g$ and $\delta T$ decouple. For any other
space-time, a perturbation $\delta T$ of one type leads to multiple
types of perturbations in $\delta g$. This feature should be useful in
suggesting more general cosmological models~\cite{anisotropy}, which
agree with the recently reported observational deviations from the
homogeneous, isotropic model in cosmology.

\begin{acknowledgments}

The work of RVB was supported by DOE grant number DE-FG02-91ER40661
and that of TWK by DOE grant number DE-FG05-85ER40226.

\end{acknowledgments}

\end{document}